# Investigation of Noise Reduction and SNR Enhancement in Search Coil Magnetometers at Low Frequencies

Abolghasem Nourmohammadi[1], S. Mohammad Hassan Feiz[2], Mohammad Hassan Asteraki[2]

*Abstract*— Here, a generalized induction coil sensor model (more generalized than other models) has been considered, and the equivalent magnetic field of the coil's thermal noise and the sensor's signal to noise ratio (*SNR*) are calculated theoretically based on the dimensions and geometry of the coil winding and its core. Our calculations indicate that the equivalent magnetic field of the thermal noise can be minimized by the coil to core weight ratio. Moreover, it is found that the sensor's *SNR* can be maximized with only a special value of core aspect ratio (length to diameter of core ratio). The calculation results here exhibit good agreement with the experimentally measured noise data.

## I. INTRODUCTION

SOME of the most important and effective instruments for measuring variable magnetic fields are induction coil sensors (search coils) which cover wide frequency ranges from $10^{-3}$ to $10^6$ Hz. Search coil magnetometers consist of a multiturn loop with a core material of high permeability, which usually increases the efficiency and response of these sensors. Compared with other magnetometers, low power consumption of these magnetometers makes them suitable for many magnetic detection applications such as magnetic anomaly detection, magnetotellurics, traffic flow detection, spacecraft and satellite applications [1-6]. However, the response of the induction coil sensors is limited by different sources of noise such as thermal and magnetic noises. Thermal noise depends on resistance of the coil winding, medium temperature and magnetic field frequency bandwidth [6].

Many authors have already attempted minimizing of the thermal noise amplitudes in search coil magnetometers [7-9]. Lukoschus has considered the effect of the geometric parameters on the equivalent magnetic field of thermal noise in order to obtain the optimum sensor geometry for thermal noise minimization [7]. However, in his optimization theory, he has introduced parameter A in equation (30) of his paper and has assumed that it is constant (see Appendix A, this article), whereas it is quite clear, A could not be a constant and depends on variable $\beta$ which is defined later in this article (Appendix A and section IIA). Thus, holding A as a constant is not a reasonable calculation. For this reason, Lukoschus has found the optimum values for the coil to core weight ratios, namely, q<1.

Here, first, we selected parameter $\beta$ as a variable one in order to improve Lukoschus' approach. Our calculation resulted in an optimum value of q greater than two. Second, optimum values were found for other involved parameters.

Also, A. Grosz and E. Paperno have optimized diameters of the ferromagnetic core and the winding wire in their search coil in order to minimize thermal noise by using some assumptions and constraints, such as holding constant the sensor's volume and aspect ratio [9]. In our optimization calculations, all involved parameters were considered without any assumptions and constraints (Fig. 1).

Finally, it was found that our calculation results agree with the experiment.

## II. ANALYTICAL METHODS

Search coil sensors work based on the Faraday's law of induction. According to this law, *e.m.f* induced in a coil subjected to a varying magnetic field is given by:

$$V = -n \cdot \frac{d\Phi}{dt} = -\mu_0 n A \frac{dH}{dt} \qquad (1)$$

Here, $\Phi$ is the magnetic flux passing through the sensor's coil, A is the average turn area of the coil, n is number of turns, $\mu_0$ is vacuum permeability constant, H is applied magnetic field and V is induced potential (sensor's output). The output of these sensors can be effectively increased using a core made of a soft magnetic material with high relative permeability ($\mu_r$); see (2):

$$V = -\mu_0 \mu_r n A \frac{dH}{dt} \qquad (2)$$

[1] A. Nourmohammadi is at the Department of Nanotechnology, Faculty of Advanced Science and Technologies, University of Isfahan, 81746-73441, Isfahan, Iran (e-mail: a.nourmohammadi@sci.ui.ac.ir).

[2] S. M. H. Feiz and M. H. Asteraki are at the Department of Physics, Faculty of Science, University of Isfahan, 81746-73441, Isfahan, Iran

Thus, for a sinusoidal varying magnetic field, $H = H_a \sin 2\pi ft$, the maximum amplitude of variation will be:

$$V_{max} = 2\pi f \mu_0 \mu_r nAH_a \tag{3}$$

Where, $f$ is frequency of the applied magnetic field and $H_a$ is its amplitude. The sensitivity constant of the search coils, $S$, can be defined by normalizing the maximum voltage, $V_{max}$, with respect to $H$ and $f$, as follows:

$$S = \frac{V_{max}}{H_a \cdot f} = 2\pi \mu_0 \mu_r nA \tag{4}$$

From (4), it is clear that the sensitivity constant, $S$, increases linearly with an increase in the relative permeability of the core, $\mu_r$, and number of turns, $n$. However, the improving effect of these parameters is limited because raising theses parameters may amplify thermal and magnetic noises. Magnetic noises are due to the ferromagnetic core and thermal noise depends on the electrical resistance of the coil, $R$. Therefore, increasing number of turns, $n$, can increase the thermal noise because it raises the resistivity.

Thermal noise voltage, $V_n$, depends on the electrical resistance of the coil, $R$, the medium temperature, $T$ in Kelvin, and magnetic field frequency bandwidth, $\Delta f$. It can be written as[10]:

$$V_n = \sqrt{4k_B TR \cdot \Delta f} \tag{5}$$

Where $k_B$ is the Boltzmann constant.

Moreover, using (4), $H_a = \frac{V_{max}}{S \cdot f}$, the equivalent magnetic field for thermal noise, $H_e$, can be calculated by the following relation:

$$H_e^2 = \frac{4k_B TR \cdot \Delta f}{S^2 \cdot f^2} \tag{6}$$

Fabrication of a high response induction coil sensor necessitates sophisticated calculations of the possible noise amplitudes, $H_e$, and optimization of SNR. These topics are considered here, as follows:

A. *Calculation of Minimizing the Equivalent Magnetic Field for Thermal Noise*

Equation (6) shows that the equivalent magnetic field for thermal noise, $H_e$, is proportional to the electrical resistance of the coil, $R$ which, in turn, depends on the geometric parameters of the coil. Hence, the correlation between $H_e^2$ and geometric parameters of coils was found here in order to minimize $H_e^2$.

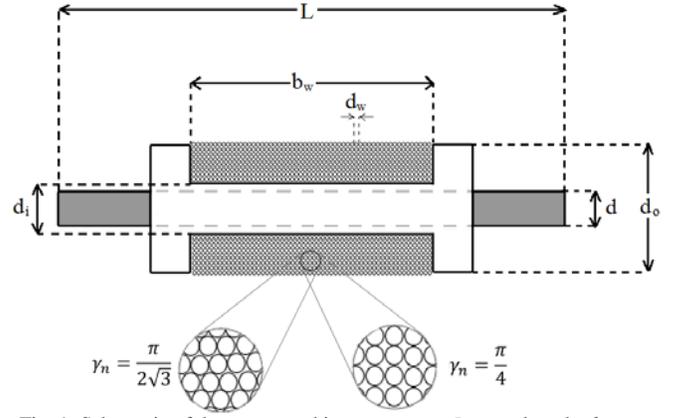

Fig. 1. Schematic of the sensor and its parameters, $L$: core length, $d$: core diameter, $b_w$: coil length, $d_o$: outer diameter of the coil, $d_i$: inner diameter of the coil, $\gamma_n$: winding filling factor, $d_w$: diameter of the coil wire.

Fig. 1 shows a search coil sensor schematically [6]. According to this figure, the total length of the winding wire is $n \times l_m$ and coil resistance, $R$, can be obtained by the formula:

$$R = \rho_w \frac{n \cdot l_m}{d_w^2} \tag{7}$$

Where $\rho_w$, $n$, $l_m$ and $d_w$ are resistivity of coil, number of turns, mean length per turn and diameter of the coil wire, respectively (Fig. 1). Besides, the weight of the coil, $W_w$, is given by:

$$W_W = \frac{\pi}{4} d_w^2 \delta_w n l_m \tag{8}$$

Where, $\delta_w$ is the density of coil wire. According to Fig. 1., number of turns per layer and number of winding layers will be $\frac{b_w}{d_w}$ and $\frac{d_o - d_i}{2d_w}$, respectively. Here, it is assumed that the thickness of the insulation layer of the wire is negligible. Therefore, in an ideal case of a regular orthogonal winding, number of turns can be expressed by:

$$n = \frac{b_w}{d_w} \cdot \frac{d_o - d_i}{2d_w} = \beta m\alpha d^2 \frac{z-1}{2d_w^2} \gamma_n \tag{9}$$

Where $\beta = \frac{b_w}{L}$, $z = \frac{d_o}{d_i}$, $m = $(core aspect ratio)$ = \frac{L}{d}$, $\alpha = \frac{d_i}{d}$ and $\gamma_n$ is the coil's filling factor (Fig. 1.) [6]. Consequently:

$$d_w^2 = \beta m\alpha d^2 \frac{z-1}{2n} \gamma_n \tag{10}$$

By combining (7), (8) and (10) we obtain:

$$R = \left(\frac{4}{\pi}\right)^2 \frac{\rho_w W_w 4n^2}{(\alpha\beta m\gamma_n)^2 d^4 (z-1)^2} \quad (11)$$

The weight of the core, $W_c$, is given by its volume and its density, $\delta_c$, by the formula:

$$W_c = \frac{\pi}{4} m d^3 \delta_c \quad (12)$$

By putting $\frac{W_w}{W_c} = q$ and $W_w + W_c = W_s$ in (12), it can be written:

$$d = \left(\frac{4W_s}{\pi m \delta_c (1+q)}\right)^{\frac{1}{3}} \quad (13)$$

By considering these parameters, if we define $A$ and $B$ as: $A = 4k_B T \cdot \Delta f / f^2$ and $B = \rho_w \delta_c^{\frac{8}{3}} / (2\pi)^{\frac{4}{3}} \delta_w \gamma_n^2 \alpha^8$, $H_e$ can be calculated by:

$$H_e^2 = AB \frac{m^{\frac{2}{3}}(1+q)^{\frac{11}{3}}}{q\beta^2(z-1)^2} \frac{1}{W_s^{\frac{5}{3}}} \quad (14)$$

This equation expresses dependence of $H_e$ on the geometric parameters of the coil $m$, $q$, $\beta$ and $z$ as already defined. Therefore, in order to determine the optimum values of these parameters, we try to find a relation between the geometric parameters of the coil viz., $m$, $q$, $\beta$ and $z$. To do this, according to Fig.1 and (10), we write a second relation for $W_w$ by the formula:

$$W_w = n \cdot \pi \frac{d_o + d_i}{2} \cdot d_w^2 \cdot \delta_w = \frac{\pi}{4}(z^2-1)\alpha^2 \beta m d^3 \gamma_n \delta_w \quad (15)$$

Dividing (15) by (12) can give the equation containing these geometric parameters (16):

$$q - (z^2-1)\beta\alpha^2 \frac{\delta_w}{\delta_c} = 0 \quad (16)$$

By using (16), $H_e$ can be minimized for an optimized $q$, namely $q_{op}$. The optimum values of $q$ are shown in Fig. 2, which depend on the core density, $\delta_c$, and parameter $\alpha$. It is observed in Fig. 2 that all the values of $q_{op}$ calculated here are notably higher than the $q_{op}$ which has been already reported by Lukoschus [7].

Accordingly, optimum value of z can be calculated from (16).

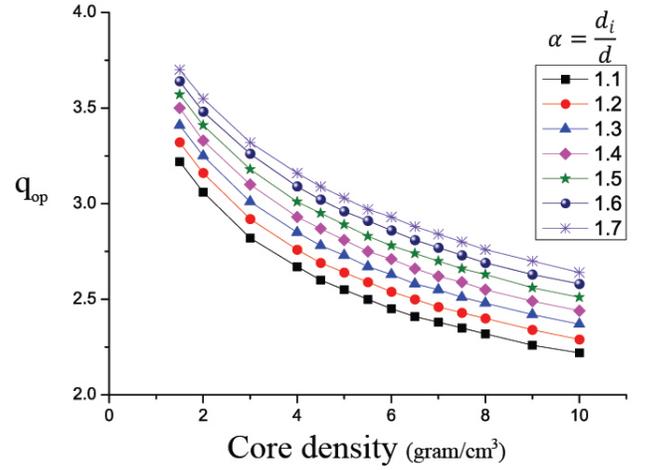

Fig. 2. (Color online) Dependence of the optimum value of $q$ on the core density ($\delta_c$) for different values of the parameter $\alpha$

B.     *Calculation of Maximizing Signal to Noise Ratio (SNR)*

Regarding $\mu_r$ is related to the core material permeability, it is usually larger than the effective permeability of the core (18), because core magnetization produces a demagnetizing field inside the core [6]. Hence, the magnetic field inside the core is $H_i = H - N.M$ where $M = (\mu_r - 1)H_i$ is the magnetization of the core and $N$ is the demagnetizing factor. The resultant field inside the core will be:

$$H_i = \frac{H}{1 + N(\mu_r - 1)} \quad (17)$$

Regarding (17), an effective permeability $\mu_c$ could be defined for the core material as follows:

$$\mu_c = \frac{\mu_r}{1 + N(\mu_r - 1)} \quad (18)$$

Hence, the parameter $\mu_r$ was replaced by $\mu_c$ in our all calculations.

The demagnetizing factor depends only on the geometric shape of the core. For an ellipsoidal core with an ideal permeability ($\mu_r \to \infty$) and magnetization parallel to its major axis, the demagnetizing factor, $N_{ell\infty}$, depends on aspect ratio ($m$) of the core and is calculated exactly using the results of articles [11-12]:

$$N_{ell\infty} = \frac{1}{m^2 - 1}\left[\frac{m}{\sqrt{m^2-1}} Ln\left(m + \sqrt{m^2-1}\right) - 1\right] \quad (19)$$

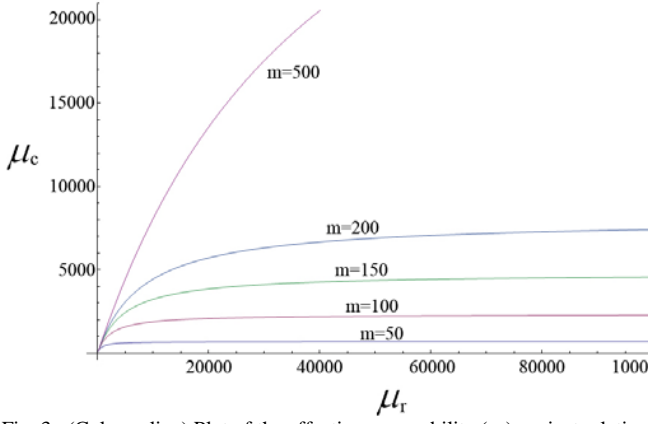

Fig. 3. (Color online) Plot of the effective permeability ($\mu_c$) against relative permeability ($\mu_r$) for different value of aspect ratio ($m$).

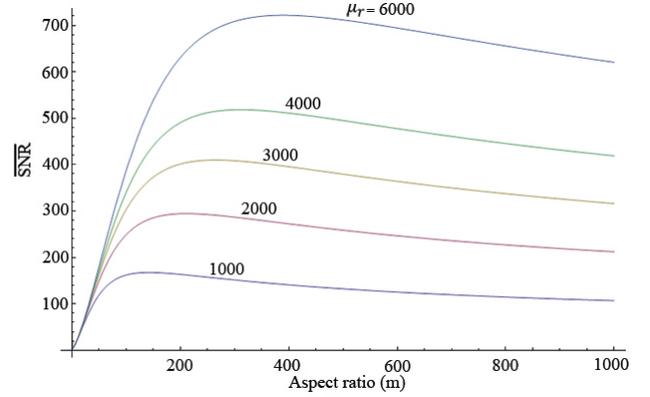

Fig. 4. (Color online) Variation of *SNR* with the core aspect ratio for a given core permeability, $\mu_r$=(1000-6000)

Bozorth has already reported that effective permeability of the core reaches to a saturation case for different values of aspect ratio, $m$ [12]. If the aspect ratio is low (e.g., 10 or lower) the effective core permeability $\mu_c$ won't be proportional to the relative permeability $\mu_r$ (Fig. 3), so that, sensor's output (signal) can't be improved through $\mu_r$. According to Fig. 3, $\mu_c$ tends to $\mu_r$ when $m$ is sufficiently large. However, our calculations in (12) clearly show that $H_e$ increases with an increase in the aspect ratio of the coil, to the power of 2/3. Hence, in the current research study, we calculated signal to noise ratio (SNR) from (3) and (5) as follows:

$$SNR = \frac{V_{max}}{V_n} = \frac{\gamma_n \pi^{\frac{2}{3}}}{2^{\frac{1}{3}} \delta_c^{\frac{4}{3}} \sqrt{A}} \cdot \mu_0 \mu_c H_a \frac{\beta(z-1) W_s^{\frac{5}{6}}}{m^{\frac{1}{3}} q^{\frac{1}{2}} (1+q)^{\frac{5}{6}}} \quad (20)$$

Putting $\mu_c$ from (18) into *SNR* and using (19), the dependence of the signal to noise ratio (*SNR*) on core aspect ratio can be obtained. To get a more general expression, the following normalization is introduced:

$$SNR = \overline{SNR} \frac{\gamma_n \pi^{\frac{2}{3}}}{2^{\frac{1}{3}} \delta_c^{\frac{4}{3}} \sqrt{A}} \cdot \mu_0 H_a \frac{\beta(z-1) W_s^{\frac{5}{6}}}{q^{\frac{1}{2}} (1+q)^{\frac{5}{6}}} \quad (21)$$

Where, by using (19):

$$\overline{SNR} = \frac{\mu_r}{m^{\frac{1}{3}}} \cdot \frac{1}{\left\{1 + \frac{1}{m^2-1}\left[\frac{m}{\sqrt{m^2-1}} Ln\left(m+\sqrt{m^2-1}\right) - 1\right](\mu_r - 1)\right\}} \quad (22)$$

Fig. 4 shows the plot of $\overline{SNR}$ against the core aspect ratio for a given value of $\mu_r$, and Fig. 5 shows dependence of the optimized aspect ratios of Fig. 4 on the core permeability, as obtained by calculation.

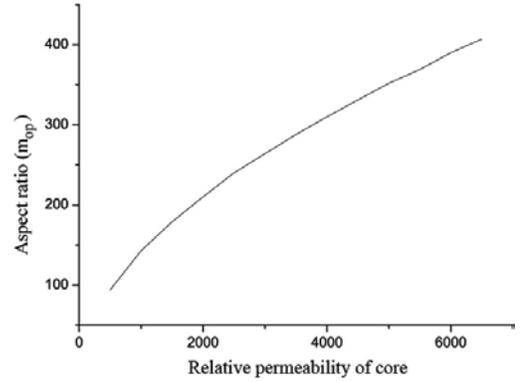

Fig. 5. Dependence of the optimum aspect ratio ($m_{op}$) on core permeability.

The following important and interesting results must be noted:

a) For all m, an increase in $\mu_r$ causes an increase in $\overline{SNR}$ (Fig. 4).
b) At a given value of $\mu_r$, $\overline{SNR}$ increases with an increase in $m$, and for a particular value $m$, called $m_{op}$, its value is maximum. Beyond this value, if the aspect ratio $m$ increases, $\overline{SNR}$ will decrease (Fig. 4).
c) The optimum aspect ratio, $m_{op}$, at which maximum $\overline{SNR}$ takes place, increases with an increase in $\mu_r$ (Fig. 5).

III. EXPERIMENT

By considering the optimum value of the geometric parameters achieved from our calculations, a model search coil magnetometer was fabricated and studied. The model magnetometer is shown in Fig. 6 and its parameters are listed in Table I. In order to detect the output, the induced voltage, a low noise operational amplifier AD620 was used and appropriate electronic circuits were designed and produced for this amplifier.

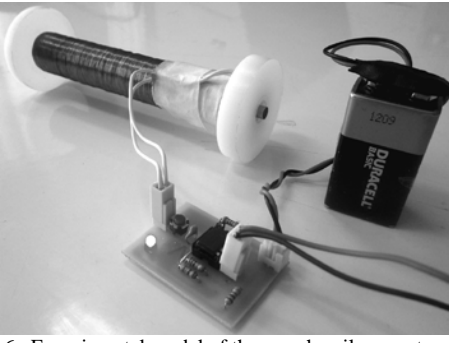

Fig. 6. Experimental model of the search coil magnetometer.

TABLE I
SPECIFICATIONS OF SEARCH COIL MAGNETOMETER

| Symbol | Quantity | Value |
| --- | --- | --- |
| $\beta$ | $b_w/L$ | 1 |
| $z$ | $d_o/d_i$ | 2 |
| $q$ | $W_w/W_c$ | 2.7 |
| $L$ | Core length | 17 cm |
| $d_w$ | Diameter of the coil wire | 0.1 mm |
| $\mu_r$ | Relative permeability of core | 4000 |
| $n$ | Number of turns | 90000 |
| $\alpha$ | $d_i/d$ | 1.4 |
| $d$ | Core diameter | 5 mm |
| $W_s$ | Total weight | 200 gram |
| $H_e$ | Equivalent magnetic field of Thermal noise at 1Hz | Theoretical  : 11.88 pT/√Hz<br>Experimental : 12.5 pT/√Hz |

As is clear from the Table I, the experimentally measured equivalent magnetic field for thermal noise (12.5 pT/√Hz, at 1 Hz) is very close to the theoretically calculated one (11.88 pT/√Hz).

IV. CONCLUSION

In this paper, the basic principle and structure of the search coil with a ferromagnetic core were elaborated on, and a generalized model of search coil was introduced. The equivalent magnetic field of the thermal noise, $H_e$, was studied theoretically and experimentally. The effect of dimensions and geometry of coil winding and its core on $H_e$ were investigated. It was shown that $H_e$ is minimized by optimizing the coil to core weight ratio, $q_{op}$. Moreover, it was shown that the optimum coil to core weight ratio can be larger than values previously calculated by the other authors. It was found that optimizing signal to noise ratio leads to an optimized core aspect ratio, which can be increased by an increase in $\mu_r$. Finally it was found that only given values of core aspect ratio (not infinitive) are capable of maximizing signal to noise ratio. Also, in this study, we did not apply any constraints and assumptions on the geometrical parameters applied previously by other authors. Accordingly, our optimization calculations are more comprehensive compared with other studies cited before and are more advantageous in terms of generalizability of the results.

**Appendix A:**

Lukoschus calculated the equivalent magnetic field of the thermal noise considering sensor's geometry parameters. His calculations resulted in equation (36) of his paper, rewritten here as (A.1):

$$\bar{H}_R^2 = \frac{4kTA.\Delta f}{D.f^2} \cdot \frac{Y(1+q)}{W_s^{5/3}} \quad \text{(A.1)},$$

It is observed in (A.1) that $\bar{H}_R^2$ is proportional to parameter A which depends on $\beta^{-\frac{8}{5}}$, according to equation (30) of his paper (A.2).

$$A = \frac{2^{14/3} \rho_w m^{2/3} \delta_c^{5/3}}{\pi^{4/3} \mu_0 \mu_c \delta_w \alpha^2 \beta^{8/5}} \quad \text{(A.2)},$$

Then, Lukoschus removed this parameter from the optimization procedure by introducing the normalized equivalent magnetic field of thermal noise in the equation (36) of his paper (A.3).

$$\bar{H}_{RN}^2 = \frac{\bar{H}_R^2 . D.f^2}{4kTA.\Delta f} = \frac{1}{W_s^{5/3}} \cdot \frac{q(1+q)^{5/3}}{\left[(1+Bq)^{1/2}-1\right]^2} \quad \text{(A.3)},$$

where,

$$B = \frac{\delta_c}{\alpha^2 \beta \gamma_n \delta_w} \quad \text{(A.4)}$$

Hence, in Lukoschus' calculations [7], the effect of the variations of parameter A has not been considered at all, while A depends on the variable $\beta$ (A.5).

$$\beta = \frac{b_w}{L} \quad \text{(A.5)}$$

Therefore, parameter A has to be considered in an optimization procedure based on sensor's geometry parameters, because it is a geometric variable. In the calculations here, the effect of this parameter has been taken into account. Therefore, it is clear that the calculations here are more generalized than the Lukoschus' calculations.